# A Methodology for Studying VANET Performance with Practical Vehicle Distribution in Urban Environment

Ivan Wang-Hei Ho[*], Kin K. Leung[†], John W. Polak[‡]

[*]Department of Electronic and Information Engineering, The Hong Kong Polytechnic University
[†]Department of Electrical and Electronic Engineering, Imperial College London
[‡]Centre for Transport Studies, Imperial College London
ivanwh.ho@polyu.edu.hk, kin.leung@imperial.ac.uk, j.polak@imperial.ac.uk

**Abstract** – In a Vehicular Ad-hoc Network (VANET), the amount of interference from neighboring nodes to a communication link is governed by the vehicle density dynamics in vicinity and transmission probabilities of terminals. It is obvious that vehicles are distributed non-homogeneously along a road segment due to traffic controls and speed limits at different portions of the road. The common assumption of homogeneous node distribution in the network in most of the previous work in mobile ad-hoc networks thus appears to be inappropriate in VANETs. In light of the inadequacy, we present in this paper an original methodology to study the performance of VANETs with practical vehicle distribution in urban environment. Specifically, we introduce the stochastic traffic model to characterize the general vehicular traffic flow as well as the randomness of individual vehicles, from which we can acquire the mean dynamics and the probability distribution of vehicular density. As illustrative examples, we demonstrate how the density knowledge from the stochastic traffic model can be utilized to derive the throughput and progress performance of three routing strategies in different channel access protocols. We confirm the accuracy of the analytical results through extensive simulations. Our results demonstrate the applicability of the proposed methodology on modeling protocol performance, and shed insight into the performance analysis of other transmission protocols and network configurations in vehicular networks. Furthermore, we illustrate that the optimal transmission probability for optimized network performance can be obtained as a function of the location space from our results. Such information can be computed by road-side nodes and then broadcasted to road users for optimized multi-hop packet transmission in the communication network.

**Index Terms** – Vehicular Ad-hoc Network, Stochastic Traffic Model, Inhomogeneous Node Distribution, Throughput and Progress, Optimal Transmission Probability.

I. INTRODUCTION

In a Vehicular Ad-hoc Network (VANET), the amount of interference to a communication link depends on the number of concurrent transmissions in vicinity, which is governed by the vehicular density dynamics, the transmission probabilities of terminals, the channel access protocol and routing strategy used.

There are a number of previous studies on analyzing capacity or throughput and forward progress (in unit distance with respect to the direction of the final destination) in Mobile Ad-hoc Networks (MANETs). For instance, [1-3] explore how network capacity scales with the number of nodes in the network. [4] investigates the optimal transmission radii for maximized expected forward progress in multi-hop packet radio network, while [5] analyzes the throughput and progress performance of several transmission strategies with transmission radius control. For VANET research, [6,7] investigate the capacity of VANETs and its scalability in highway and urban grid structures.

However, when considering these previous studies as a whole, it appears that all of them primarily assume that node density is homogeneous throughout the whole network, and lack of a general approach for handling mobile nodes that spatially distribute in a heterogeneous manner. Apparently, these existing studies are not applicable to VANETs in urban road networks, specifically with traffic signals and stops, since we expect car density at road junctions (where traffic signals are located) behaves quite differently from that at the middle of a road segment. Given the inhomogeneous distribution of vehicles in urban road networks, this paper proposes a practical methodology to characterize and optimize protocol performance in VANETs.

The spatial distribution of nodes in a VANET is governed by the space and time dynamics of moving vehicles. To capture such dynamics and thus the heterogeneity of the density distribution of vehicles, we utilize the stochastic traffic model in [8,9] for modeling vehicular traffic in generic urban road systems. The stochastic traffic model uses a fluid model to characterize the space and time dynamics of vehicle movements, which is driven by a velocity profile as a function of space and time. In real practice, empirical velocity measurements from inductive loop detectors and navigation systems can serve as inputs to the model. The average density profile, again as a function of space and time, is readily computable from the conservation equations in the fluid model. The randomness of individual vehicle is captured by a stochastic model. The actual number of vehicles in a given road section at a certain time instance has Poisson distribution according to previous results in [10,11] given that the arrival of vehicles follow a non-homogeneous Poisson process. It is also validated in [9,12] that the Poisson distributional result can be treated as an approximation even when we consider interactions between vehicles.

In this paper, we illustrate how the vehicular density dynamics acquired from the stochastic traffic model can be applied to analyze the throughput and progress performance of packet transmission in a VANET with more practical, inhomogeneously distributed nodes along a generic urban road segment. Different channel access protocols and routing strategies have different throughput and progress performance with regard to certain spatial distribution of mobile nodes. As illustrative examples, we attempt to consider two protocols, specifically slotted ALOHA and CSMA, and three basic routing strategies, that vary in the way that the packets are routed or the transmission ranges are controlled, so that we can gain insights for the analysis of other communication protocols, routing strategies, and network configurations in VANETs. Through the vehicular density dynamics computed from the stochastic traffic model, we determine the distribution of a node's location on the urban route, and derive the probability that such a node is being interfered by its neighbors in its vicinity. With this interference factor, we model the local throughput and average progress of the routing strategies for different transmission protocols.

From our analytical results, we can identify the optimal transmission probability for maximized expected progress as a function of the location space. In practice, such information can be computed by road-side infrastructure nodes based on gathered empirical velocity and vehicular flow measurements, and then broadcasted to vehicles that enter the road segment so that maximized packet propagation rate can be achieved, which is important for transport functions such as traffic information exchange and accident warning in VANETs.

The rest of the paper is organized as follows. Section II provides the background information of the stochastic traffic model. Section III introduces the network models including the channel access protocols, routing strategies, and interference model considered in this paper. In Section IV, we determine the probability that a random node in the network is being interfered by its neighbors, followed by deriving the local throughput and average progress of slotted ALOHA and CSMA protocols of three routing strategies in Section V. Section VI provides simulation results to validate the analytical model, and identifies the optimal transmission probability. Finally, Section VII concludes the paper.

## II. STOCHASTIC TRAFFIC MODEL

In this section, we define the system model for the analysis in this paper, and provide background information of the stochastic traffic model [8,9] that captures the vehicular density dynamics in generic urban routes.

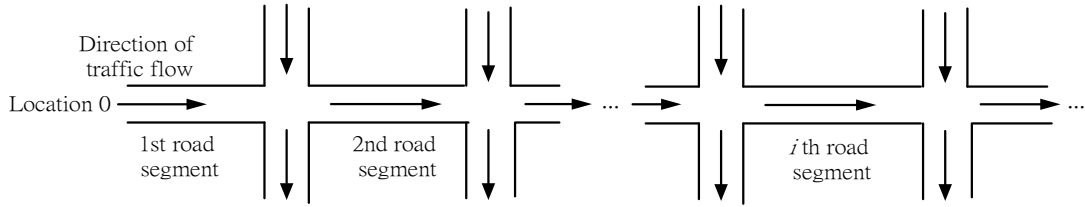

Figure 1. The road configuration considered in this paper.

We consider traffic in a one-way, single-lane, semi-infinite urban road (or route) as shown in Figure 1. Although the road is fed with traffic from adjacent streets, the one-way road under consideration is the one running from the left to the right in the figure. More complicated road topology can be represented by superposing multiple versions of urban routes. Let our location space to be the interval $[0, \infty)$, the boundary point 0 is the spatial origin, and it marks the starting point of the road. The arrival process $\{A(t) \mid -\infty < t < \infty\}$ counts the number of arrivals to the first segment of the route up to time $t$, which we assume is finite with probability 1, and is characterized by a non-negative and integrable external arrival rate function $\alpha(t)$. Furthermore, the route consists of a number of road segments indexed by $i = 1, 2, 3,\ldots$, where vehicles can leave and join the route at the junctions of road segments.

## A. Deterministic Fluid Model

The fluid model is a kind of continuum traffic flow models, which reduces laws of traffic to a partial differential equation (PDE) that may be studied as adequately as other physical phenomena that are also governed by PDE's.

The major difference between our fluid model and other continuum models is that we model vehicle motions with a velocity profile, vehicles at location $x$ and time $t$ move forward the route according to a velocity field $v(x, t)$. Stopping or slowing down of vehicles at road junctions or traffic signals can be reflected and modeled by the velocity profile. However, continuum model alone is unable to capture traffic instability and the randomness of individual vehicle, therefore, we couple the fluid model with the stochastic model as a remedy.

Let us first describe the fluid dynamic conservation equations and corresponding notations that hold for the general systems. Let $N(x, t)$ be the number of vehicles in location $(0, x]$ at time $t$, and $n(x, t)$ be the density of vehicles in location $(0, x]$ at time $t$. Let $Q(x, t)$ be the number of vehicles moving past position $x$ before time $t$, and $q(x, t)$ be the flow rate. Thus,

$$n(x,t) = \frac{\partial N(x,t)}{\partial x} \quad \text{and} \quad q(x,t) = \frac{\partial Q(x,t)}{\partial t}. \tag{1}$$

Let $C^+(x, t)$ and $C^-(x, t)$ be the number of vehicles arriving to and departing from the route in location $(0, x]$ during time interval $(-\infty, t]$, respectively. Then the associated rate densities are respectively

$$c^+(x,t) = \frac{\partial^2 C^+(x,t)}{\partial x \partial t} \quad \text{and} \quad c^-(x,t) = \frac{\partial^2 C^-(x,t)}{\partial x \partial t}. \tag{2}$$

Assuming all traffic moves only from left to right down the positive real line, then the four variables $N, Q, C^+, C^-$ satisfy the following conservation relation:

$$C^+(x,t) = N(x,t) + Q(x,t) + C^-(x,t). \tag{3}$$

By applying the operator $\partial^2/(\partial x \partial t)$ to (3), we have the partial differential equation

$$\frac{\partial n(x,t)}{\partial t} + \frac{\partial q(x,t)}{\partial x} = c^+(x,t) - c^-(x,t). \tag{4}$$

According to traffic flow theory [13], we have

$$q(x,t) = n(x,t)v(x,t). \tag{5}$$

By substituting (5) into (4), we have

$$\frac{\partial n(x,t)}{\partial t} + \frac{\partial [n(x,t)v(x,t)]}{\partial x} = c^+(x,t) - c^-(x,t). \tag{6}$$

The resulting partial differential equation (6) is a one-dimensional version of the generalized conservation law for fluid motion [14]. This equation governs the mean behavior of any stochastic traffic model.

We assume that vehicles arrive at the first route segment according to an external arrival rate function $\alpha(t)$. Let us use $\xi_i(t)$ to denote the external arrival rate of vehicles at the $i$-th junction at time $t$. Then we have

$$c^+(x,t) = \alpha(t)\delta(x) + \sum_i \xi_i(t)\delta(x - x_i). \tag{7}$$

As for vehicles leaving the route, we use $\rho_i(t)$ to denote the fraction of vehicles departing when they pass by the $i$-th junction at time $t$. If these departing vehicles leave at the same velocity as they move forward along the route, then

$$c^-(x,t) = \beta(x,t)n(x,t),$$

$$\text{where } \beta(x,t) = v(x,t)\sum_i \rho_i(t)\delta(x - x_i). \tag{8}$$

## B. Stochastic Model

In contrast to the deterministic fluid model, the stochastic model captures the stochastic fluctuations of the quantities of interest. When the two models are coupled with each other to form the stochastic traffic model, the solutions from the PDE's describe the expected number of vehicles, and the actual number of vehicles is captured by the additional distribution information from the stochastic model.

From now on, the densities $n(x, t)$ and $q(x, t)$ are defined as the partial derivatives of expected values, that is,

$$n(x,t) = \frac{\partial E[N(x,t)]}{\partial x} \quad \text{and} \quad q(x,t) = \frac{\partial E[Q(x,t)]}{\partial t}.$$

Similarly, the rate densities $c^+(x, t)$ and $c^-(x, t)$ are the second partial derivatives of expected values, that is,

$$c^+(x,t) = \frac{\partial^2 E[C^+(x,t)]}{\partial x \partial t} \quad \text{and} \quad c^-(x,t) = \frac{\partial^2 E[C^-(x,t)]}{\partial x \partial t}.$$

The general stochastic model can be of any distributions depending on the arrival process of vehicles, and the equations in the fluid model continue to hold regardless the distribution of the stochastic model. In this paper, we specifically consider the Poisson arrival location model (PALM) [10,11]. Again, the fluid dynamic model is not dependent on the Poisson assumption; they hold as long as the arrival process $A$ is an arbitrary point process with time-dependent arrival-rate function $\alpha$.

With PALM, the arrival process $\{A(t) \mid -\infty < t < \infty\}$ for vehicles to arrive at the first road segment of the route is a *non-homogeneous Poisson process* with non-negative and integrable external arrival rate function $\alpha(t)$. That is, the number of arrivals in the interval $(t_1, t_2]$ is Poisson with mean $\int_{t_1}^{t_2} \alpha(s) ds$.

According to [10,11], we can construct $N(x, t)$, the random number of vehicles within the range $(0, x]$ at time $t$, via stochastic integration starting with the Poisson process $A$, where $A(t)$ counts the number of vehicles arriving to the road segment up to time $t$.

$$N(x,t) = \int_{\sigma(x,t)}^{t} 1_{\{L_s(t) \in (0,x]\}} dA(s) = \sum_{n=A(\sigma(x,t))}^{A(t)} 1_{\{L_{\hat{A}_n}(t) \in (0,x]\}}. \tag{9}$$

where $\hat{A}_n$ is the $n$th jump time of $A$, counting backward from time $t$. $1_B$ is an indicator function such that it returns 1 if $B$ is true and 0 otherwise. $L_s(t)$ is the location process, which specifies the position of the vehicle on the road segment at time $t$ that arrived at time $s$. Let $\sigma(x, t)$ denote the route entrance time for a vehicle to be in position *x* at time *t*. For vehicles that arrive to the route before $\sigma(x, t)$, it will be past position *x* by time *t*. On the other hand, for vehicles arrive after $\sigma(x, t)$, it will be still in position *x* at time *t*. For all real *t*, [8] is a Poisson process with

$$E[N(x,t)] = \int_{\sigma(x,t)}^{t} \alpha(s)ds. \qquad (10)$$

As long as we model the traffic flow through a deterministic velocity field as a function of space and time such that vehicles do not interact with each other, the Poisson distributional conclusion (the number of vehicles in a certain road segment is a Poisson process) remains valid. We have also demonstrated in [8,9] that $N(x, t)$ can be approximated with a Poisson distribution even when we introduce vehicle interactions through a density dependent velocity profile, given that the arrival rate of vehicle is not too high (< 30 cars/min). For real-world validations of the Poisson distributional result, the reader to referred to Section 4.6 in [12], in which flow and occupancy data collected by inductive loop detectors in Central London are used.

Given that empirical velocity profile can be constructed based on data collected by navigation systems, the stochastic traffic model is a useful tool to characterize the practical density dynamics of vehicles in urban road networks. In the followings, we illustrate how the density knowledge generated can be applied to characterize and optimize the performance of packet transmission in VANETs.

## III. NETWORK MODELS

### A. Channel Access Protocols

Given the spatial distribution of mobile nodes in the network, different channel access protocols will give rise to different throughput and progress performance. As illustrative examples, we consider in this paper two conventional protocols, namely slotted ALOHA and non-persistent carrier-sense-multiple-access (CSMA) [15]. Please note that the proposed methodology is not protocol-specific, and the analytical results presented in this paper can be extended or specialized for analysis of other network protocols. For instance, results for the CSMA/CA protocol that is used in IEEE 802.11 can be derived from the CSMA results.

For slotted ALOHA, time is divided into slots of duration equal to the transmission time of a packet. In every slot, each node tries to transmit according to a Bernoulli process with parameter $p$, where $0 < p \leq 1$. That is, it is transmitting with probability $p$ and not transmitting with probability $1 - p$. It is assumed that all nodes always have packets waiting to be sent (for exchange of real-time traffic information), and a separate channel is available for acknowledgement traffic. We further assume that the system is independent from slot to slot such that whenever there is a packet waiting to be sent, this packet will be destined to any other node in the network with equal probability, no matter whether it is a new packet or retransmitted packet.

For slotted non-persistent CSMA, the constant packet transmission time is chosen as the unit of time, plus the length of a mini-slot, denoted by $\tau$, that represents the propagation delay. Hence, the length of a successful transmission period is $1 + \tau$ or $T + 1$ mini-slots, where $T = 1/\tau$, as illustrated in Figure 2. We assume that all nodes within the Carrier-Sensing Range (*CSRange*) of the transmitter recognize the transmission in one mini-slot, and they hear the transmission one mini-slot more after the completion of transmission. In every mini-slot, each node listens to the channel with probability $p'$ (except during transmission), and does not with probability $1 - p'$. If the channel is idle, it begins transmission in the same slot with probability 1. If the channel is sensed busy, it suppresses the transmission, and stops sensing the channel until the end of the current transmission.

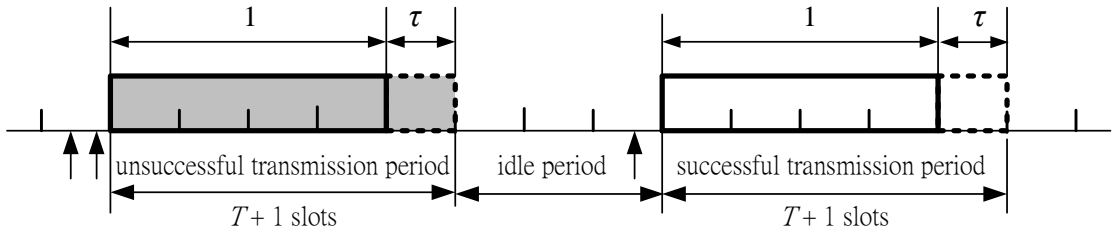

Figure 2. Transmission and idle periods of slotted non-persistent CSMA.

## B. Routing Strategies

In many applications of VANETs, cars exchange road-condition information in order to predict travel time and prevent congestions and accidents. We consider the scenario that vehicles aim to inform the following cars of the preceding road conditions for travel time prediction, and to warn following cars to slow down in case of any accidents, thus, the direction of packet transmission is opposite to that of vehicular traffic flow as depicted in Figure 3.

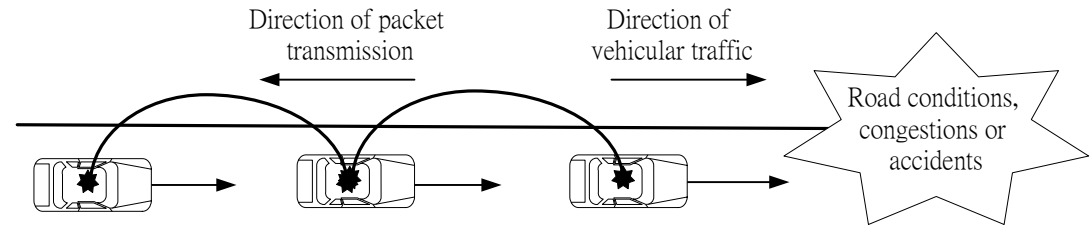

Figure 3. Packet transmission for preceding road conditions.

Each node is assumed to have a maximum communication range $R$ and know the positions of those vehicles within $R$. One simple way for a vehicle to know its position is to use Global Positioning System (GPS). We assume that this information is piggybacked on data packets. Please note that the requirement of knowing neighbors' positions is for the transmitter to determine the relay node in the routing strategies considered in this paper, the computation of protocol performance from vehicular density knowledge does not necessarily require such information. A source node will choose one of the nodes as the relay within the maximum transmission range according to one of the three routing strategies described below and transmit a packet with the identity of the relay and the identity of the final destination in the packet header. A node receiving this packet will only process the packet if it is identified as a relay, all other nodes will discard the packet.

By defining *progress* as the distance between the transmitting and receiving nodes projected onto a line drawn from the transmitter toward the final destination (or simply the distance between the transmitter and receiver for one-dimensional networks), we consider the following three basic routing strategies in this paper, similar strategies are considered in [5]:

1. *Most Progress with Fixed Radius R* (*MPR*): A node relays to the most distanced neighbor within $R$ in the backward direction to achieve the largest progress. It will use a transmission radius $R$ regardless of the position of the receiving node (i.e., there is no transmission range or power control for this strategy). The goal here is to minimize the number of hops needed for a packet to reach its destination.

2. *Most Progress with Variable Radius* (*MP*): This strategy is similar to MPR except that the transmission radius is adjusted to be the distance between the transmitting and relay nodes assuming that each node knows the positions of vehicles within its maximum transmission range. Thus, this strategy attempts to reduce the interference to some extent while maintaining the goal of obtaining the largest possible progress.

3. *Nearest with Progress* (*NP*): A node relays to the nearest neighbor within $R$ in the backward direction, and will adjust its transmit power to be just strong enough to reach the receiving node. This strategy aims to reduce the amount of interference in the network as much as possible.

As a result, the above three routing strategies differ either in the way that the packets are being routed or the transmission ranges are controlled. Note that we restrict the transmission to the backward direction (where we define the direction of vehicular traffic flow as the forward direction) in all three strategies described above. In other words, if a transmitter cannot find a receiver within the backward maximum transmission range $R$, it will not transmit in that slot. Hence, two conditions must be satisfied for a node to transmit in a slot. First, it must be in transmit mode. Second, it must be able to find a receiver in the backward direction within its maximum transmission range.

### C. Interference Model

We assume the following power-transfer relationship: $P(a, b) = cP_a/|a - b|^\gamma$, where $P(a, b)$ is the power received by node $b$ from node $a$, $P_a$ is the transmit power of node $a$, $|a - b|$ is the distance between nodes $a$ and $b$ (for brevity, we also use $a$ and $b$ to denote the nodes' positions), $\gamma > 2$ is the path-loss exponent, and $c$ is a constant. Consider the Protocol Model in [1]. For a link $(a, b)$ to transmit successfully, we require

1) Node $a$ is in transmitting mode and node $b$ is not;

2) $|a - b| \leq R$; and

3) $P_a|i - b|^\gamma > \beta P_i|a - b|^\gamma$ for every other node $i$ simultaneously transmitting, where $\beta$ is the SIR requirement.

In MPR, since all nodes use the same transmission power/radius, we define that the transmission of node $i$ will interfere with the signal reception at node $b$ from node $a$ if

$$|i - b| \leq \beta^{1/\gamma} R. \tag{11}$$

Let $R_I = \beta^{1/\gamma} R$ be the maximum interference range. We define that two nodes are *neighbors* if they are within a distance $R_I$ of each other. Hence, for the MPR strategy, a transmission from node *a* to node *b* will be successful only if there are no neighboring nodes of *b* transmit concurrently.

Let node *j* be the receiver of node *i*'s transmission. For the MP and NP strategies, since transmitters use the minimum required power to transmit, we have

$$\frac{P_a}{|a-b|^\gamma} = \frac{P_i}{|i-j|^\gamma} = \frac{Rx_{th}}{c}, \tag{12}$$

where $Rx_{th}$ is the minimum receiver threshold.

Under the protocol model, the transmission of link $(i, j)$ will interfere with that of link $(a, b)$ if

$$P_a |i-b|^\gamma \leq \beta P_i |a-b|^\gamma. \tag{13}$$

Substitute (12) into (13), we have

$$|i-b| \leq \beta^{1/\gamma} |i-j| < \beta^{1/\gamma} R. \tag{14}$$

Hence, if the distance between nodes *b* and *i* is less than $\beta^{1/\gamma} |i-j|$, we say that node *b* falls within the interference range of node *i*, and the signal reception at *b* can be interfered by the transmission from *i*. For node *i* to be able to interfere with the reception at node *b*, it is necessary that node *i* is a neighbor of node *b* (or node *i* lies within the maximum interference range $R_I$ of node *b*).

## IV. PROBABILITY OF INTERFERENCE

For analytical simplicity, we consider a specific time instance $t_0$ and cease to focus on time dynamics in this section. For this reason, the variable *t* is simply dropped from our previously defined notations. In practice, probabilities over a period of time can be obtained by taking the time-average of multiple time instances.

Consider a one-dimensional road segment of length $L$. From the fluid model, we can acquire the mean vehicular density profile $n(x)$, such that the expected number of vehicles within a region $(x_1, x_2]$ in the road segment is

$$E[N(x_1, x_2)] = \int_{x_1}^{x_2} n(x) dx. \quad (15)$$

Moreover, we can derive the probability density function (pdf)

$$f_L(x) = n(x) / E[N(L)], \quad (16)$$

where $f_L(x)\Delta x$ represents the probability that a random node in the road segment of length $L$ is located in the small region $(x, x + \Delta x]$. For notation simplicity, let us denote $E[N(x_1, x_2)]$ as $\bar{N}_{x_1}^{x_2}$ and $\int_{x_1}^{x_2} f_L(x) dx$ as $m_{x_1}^{x_2}$, where the latter is the probability that a random node in the network is located in the region $(x_1, x_2]$.

Let $C_a$ be the event that a transmitter at position $a$ can be connected with a receiving node $b$ that lies within its backward maximum transmission range. i.e., there is at least one node located in the region $(a, a - R]$. Given that the arrival of vehicles follows a non-homogeneous Poisson process, according to the Poisson property of the stochastic traffic model, we have

$$P(C_a) = 1 - \exp(-\bar{N}_{a-R}^{a}). \quad (17)$$

Let the random variable $r_a = |a - b|$ represent the distance between nodes $a$ and $b$. For the MPR and MP strategies, the cumulative distribution function (CDF) of $r_a$ under the condition that $C_a$ occurs is

$$F_{r_a}(r) = P(r_a \leq r) = \frac{P(\text{no nodes in } (a-R, a-r]) P(\text{at least one in } (a-r, a])}{P(C_a)}$$

$$= \frac{e^{-\bar{N}_{a-R}^{a-r}}(1 - e^{-\bar{N}_{a-r}^{a}})}{1 - e^{-\bar{N}_{a-R}^{a}}} = \frac{e^{-\bar{N}_{a-R}^{a-r}} - e^{-\bar{N}_{a-R}^{a}}}{1 - e^{-\bar{N}_{a-R}^{a}}}. \quad (18)$$

While for the NP strategy, the CDF of $r_a$ given that $C_a$ occurs is

$$F_{r_a}(r) = P(r_a \leq r) = \frac{1-e^{-\bar{N}^a_{a-r}}}{1-e^{-\bar{N}^a_{a-R}}}. \tag{19}$$

For a transmission from node *a* to node *b* to be successful, node *b*'s neighboring nodes must not interfere with node *b*. Let node *i* be one of node *b*'s neighbors and the random variable $s_i = |b - i|$ be the distance between nodes *b* and *i*. Given that there is a transmission from node *a* to node *b*, there exists an *excluded region* such that no node exists. This region is a function of position *b* with respect to position *a*, and the routing strategy used.

For the MPR and MP strategies, we know that node *i* cannot be located in region $(a - R, a - r_a]$. Otherwise, node *a* will transmit to node *i* instead of node *b*. Therefore, node *i* could be in one of the following two regions as illustrated in Figure 4; namely, Region 1: $(a - r_a, a - r_a + R_I]$ and Region 2: $(a - r_a - R_I, a - R]$.

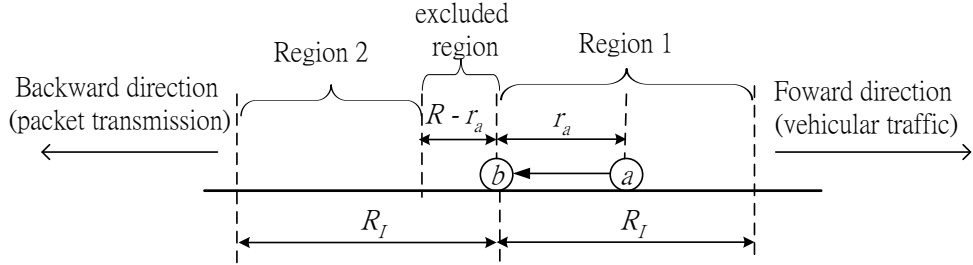

Figure 4. The regions that a neighboring node of node *b* could be located in under the MPR and MP strategies.

Similarly, for the NP strategy, we know that node *i* cannot be located in between nodes *a* and *b*. Otherwise, node *a* will transmit to node *i* instead of node *b*. Therefore, node *i* could be in one of the following two regions as illustrated in Figure 5; namely, Region 1: $(a, a - r_a + R_I]$ and Region 2: $(a - r_a - R_I, a - r_a]$.

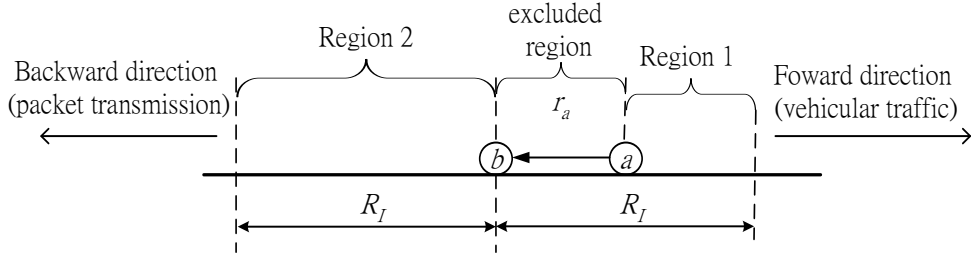

Figure 5. The regions that a neighboring node of node *b* could be located in under the NP strategy.

Let $G(g)$ denote the event that node $i$ is located in Region $g$, where $g = 1, 2$, and $I_N$ the event that node $i$ will interfere with node $b$ given that node $i$ is transmitting in the same slot. Thus, given that $r_a = r$ and the position of node $a$, we have

$$P(I_N) = \sum_{g=1}^{2} P(I_N | G(g)) \cdot P(G(g)), \tag{20}$$

where

$$P(G(g)) = \int_{G(g)} f_L(x)dx / (\int_{G(1)} f_L(x)dx + \int_{G(2)} f_L(x)dx) \tag{21}$$

$$P(I_N | G(g)) = \int_s P(s_i \leq \beta^{1/\gamma} r_i | s_i = s) f_{s_i}^{G(g)}(s)ds$$

$$= \int_s \left( \int_{s/\beta^{1/\gamma}}^{R} f_{r_i}^{G(g)}(u)du \right) f_{s_i}^{G(g)}(s)ds. \tag{22}$$

According to (14), $I_N$ happens when $s_i \leq \beta^{1/\gamma} r_i$, where $r_i = |i - j|$, such probability is represented by the inner integration in (22), $f_{s_i}^{G(g)}(s)$ and $f_{r_i}^{G(g)}(s)$ denote the respective pdf for $s_i$ and $r_i$ given that node $i$ is in Region $g$.

Let $q$ be the probability that neither transmitter $a$ nor receiver $b$ is within node $i$'s transmission range in the backward direction. For the MPR strategy, in which every node uses the same transmission range $R$, when neither nodes $a$ nor $b$ is within the backward transmission range of node $i$, node $i$ will be able to transmit with probability $1 - \exp(-\bar{N}_{i-R}^i)$ when it is in transmit mode. On the other hand, when either nodes *a* or *b* is within node *i*'s backward transmission range, node *i* can always find a receiver. Thus, given the position of node *i*, we have

$$P(s_i \leq \beta^{1/\gamma} r_i) = P(s_i \leq \beta^{1/\gamma} R) = (1-q) + q(1 - e^{-\bar{N}^i_{i-R}}) = 1 - qe^{-\bar{N}^i_{i-R}}. \tag{23}$$

Given that $r_a = r$, $s_i = s$, and the position of node *a*, if node *i* is located in Region 1, $q = m^{a-r+R_I}_{a+R} / m^{a-r+R_I}_{a-r}$ and the position of node *i* is $a - r + s$; while if it is in Region 2, $q$ simply equals to 1 since nodes *a* and *b* are not located in the backward direction of node *i*, and node *i*'s position is $a - r - s$. Hence, (20) becomes

$$P(I_N) = \frac{1}{m^{a-r+R_I}_{a-r-R_I} - m^{a-r}_{a-R}} \left[ \int_0^{R_I} (1 - qe^{-\bar{N}^{a-r+s}_{a-r+s-R}}) f_L(a-r+s) ds + \int_{R-r}^{R_I} (1 - e^{-\bar{N}^{a-r-s}_{a-r-s-R}}) f_L(a-r-s) ds \right]. \tag{24}$$

For the MP and NP strategies, where each node uses a transmission range that is just large enough to cover the receiver, when neither nodes *a* nor *b* is within the backward maximum transmission range of node *i*, node *i* will be able to transmit with probability $1 - \exp(-\bar{N}^i_{i-R})$ when it is in transmit mode. The CDF of node *i*'s transmission range, $r_i$, is $F_{r_i}(u)$ under this condition, which is given by (18) and (19) by replacing the location of node *a* with that of node *i* for the MP and NP strategies, respectively. On the other hand, when either node *a* or node *b* is within node *i*'s backward maximum transmission range, node *i* can always find a receiver. We approximate the CDF of node *i*'s transmission range as $F_{r_i}(u)$ in this case. Thus, given the position of node *i*, we have

$$P(s_i \leq \beta^{1/\gamma} r_i) \simeq \int_{s/\beta^{1/\alpha}}^{R} f^{G(g)}_{r_i}(u) du = \int_{s/\beta^{1/\gamma}}^{R} [(1-q) + q(1 - e^{-\bar{N}^i_{i-R}})] dF_{r_i}(u)$$

$$= \int_{s/\beta^{1/\gamma}}^{R} (1 - qe^{-\bar{N}^i_{i-R}}) dF_{r_i}(u). \tag{25}$$

Given that $r_a = r$, $s_i = s$, and the position of node $a$, if node $i$ is located in Region 1, $q = m_{a+R}^{a-r+R_I} / m_{a-r}^{a-r+R_I}$ for MP, and $q = m_{a+R}^{a-r+R_I} / m_a^{a-r+R_I}$ for NP, the position of node $i$ is $a - r + s$; while if it is in Regions 2, $q$ simply equals to 1 for both strategies, and node $i$'s position is $a - r - s$.

As a result, by substituting (21), (22), and (25) into (20), we have for the MP strategy

$$P(I_N) = \frac{1}{m_{a-r-R_I}^{a-r+R_I} - m_{a-R}^{a-r}} \left[ \int_0^{R_I} \left( \int_{s/\beta^{1/\gamma}}^{R} (1-qe^{-\bar{N}_{a-r+s-R}^{a-r+s}}) dF_{r_i}(u) \right) f_L(a-r+s) ds + \int_{R-r}^{R_I} \left( \int_{s/\beta^{1/\gamma}}^{R} (1-e^{-\bar{N}_{a-r-s-R}^{a-r-s}}) dF_{r_i}(u) \right) f_L(a-r-s) ds \right], \quad (26)$$

while for the NP strategy, we have

$$P(I_N) = \frac{1}{m_{a-r-R_I}^{a-r+R_I} - m_{a-r}^{a}} \left[ \int_r^{R_I} \left( \int_{s/\beta^{1/\gamma}}^{R} (1-qe^{-\bar{N}_{a-r+s-R}^{a-r+s}}) dF_{r_i}(u) \right) f_L(a-r+s) ds + \int_0^{R_I} \left( \int_{s/\beta^{1/\gamma}}^{R} (1-e^{-\bar{N}_{a-r-s-R}^{a-r-s}}) dF_{r_i}(u) \right) f_L(a-r-s) ds \right] \quad (27)$$

## V. THROUGHPUT AND PROGRESS ANALYSIS

### A. Slotted ALOHA

Let $E_i$ be the event that neighbor $i$ does not interfere with node $b$. $E_i$ will occur when node $i$ is not transmitting, or it is transmitting but it does not interfere with the signal reception at node $b$.

$$P(E_i) = (1-p) + p(1-P(I_N)) = 1 - pP(I_N). \quad (28)$$

Let $a \rightarrow b$ be the event that the transmission from node $a$ to node $b$ is successful given that node $a$ transmits to $b$, and $N_k$ be the event that node $b$ has $k$ neighbors, not including node $a$.

$$P(a \to b \mid N_k) = P(b \text{ does not transmit AND the } k \text{ neighbors do not interfere with } b)$$

$$= (1-p)\prod_{n=1}^{k} P(E_i) = (1-p)(1-pP(I_N))^k. \tag{29}$$

Again, with regard to the Poisson property of the stochastic traffic model, we have the number of vehicles within a road region be a Poisson process, since the $k$ neighbors of node $b$ cannot be located in the excluded region $(a - R, a - r_a]$ given the transmission from node $a$ to node $b$, we have

$$P(N_k \mid r_a = r) = \frac{\left(\bar{N}_{a-r-R_I}^{a-r+R_I} - \bar{N}_{a-R}^{a-r}\right)^k}{k!} e^{-\left(\bar{N}_{a-r-R_I}^{a-r+R_I} - \bar{N}_{a-R}^{a-r}\right)}. \tag{30}$$

Thus,
$$P(a \to b) = \sum_{k=0}^{\infty} P(a \to b \mid N_k) \cdot P(N_k)$$

$$= (1-p)\sum_{k=0}^{\infty}(1-pP(I_N))^k \cdot \int_0^R P(N_k \mid r_a = r)dF_{r_a}(r),$$

where $F_{r_a}(r)$ is given in (18) for the MPR and MP strategies and (19) for the NP strategy. Substitute (30) into it, we have

$$P(a \to b) = (1-p)\int_0^R \exp\left[-pP(I_N)\left(\bar{N}_{a-r-R_I}^{a-r+R_I} - \bar{N}_{a-R}^{a-r}\right)\right]dF_{r_a}(r). \tag{31}$$

To evaluate the throughput and progress of the network, we compute $\rho$, the *local throughput* (or single-hop throughput), which is the *average number of successful transmissions per slot from a node*, and $\pi$, the *average progress*, which is the *average packet propagation distance per slot from a node*. The local throughput of the transmitting node located at $a$ is given by

$$\rho_a = P(C_a) \cdot P(a \text{ transmits}) \cdot P(a \to b). \tag{32}$$

Substitute (17) and (31) into it, we have

$$\rho_a = (1 - e^{-\bar{N}_{a-R}^{a}}) p(1-p) \int_0^R e^{-pP(I_N)\left(\bar{N}_{a-r-R_I}^{a-r+R_I} - \bar{N}_{a-R}^{a-r}\right)} dF_{r_a}(r). \tag{33}$$

The average progress for a node located at $a$, $\pi_a$ can be obtained similarly by inserting the term $r$ into the integration above,

$$\pi_a = (1 - e^{-\bar{N}_{a-R}^{a}}) p(1-p) \int_0^R e^{-pP(I_N)\left(\bar{N}_{a-r-R_I}^{a-r+R_I} - \bar{N}_{a-R}^{a-r}\right)} r \, dF_{r_a}(r). \tag{34}$$

Therefore, the average local throughput and progress for the whole network in the road segment $(0, L]$ is given by the following weighted integrals.

$$\rho = \int_0^L \rho_x f_L(x) dx \quad \text{and} \quad \pi = \int_0^L \pi_x f_L(x) dx. \tag{35}$$

### B. Slotted non-persistent CSMA

To analyze the CSMA protocol, we introduce an assumption that the actual transmission occurs as a result of channel sensing are independent Bernoulli trials. That is, for every mini-slot (except during transmission), each node transmits a packet with probability $p$ (and does not with probability $1 - p$). A similar assumption was used in [4,15], and the validity of the results will be claimed by comparing the throughput values against simulation in the next section. In the following, we formulate our optimization problem with only $p$, the reader is referred to Appendix I for the determination of the channel sensing probability $p'$ in terms of $p$.

A transmission is successful when no neighbors of the receiver transmit during the transmission period $1 + \tau$ or $T + 1$ mini-slots. Potential interfering (or neighboring) nodes are located in Regions 1, 2, or 3 as illustrated in Figure 6a for the MPR and MP strategies and Figure 6b for the NP strategy. Since neighboring nodes of $b$ in Regions 1 and 2 are within the *CSRange* of transmitter $a$ (where *CSRange* > $R$), they can recognize the transmission in one mini-slot, and collision can be avoided if they do not transmit in the same slot as node $a$. While for neighbors located in Region 3, they are hidden to node $a$, it is required that they are silent throughout the vulnerable period of length $2 + \tau$ or $2T + 1$ mini-slots as shown in Figure 6c. The first $T$ mini-slots are included for preventing collisions with any ongoing transmissions, while the following $T + 1$ mini-slots are included for not being interfered with any newly started transmissions. Hence, we have

$$P(a \to b \mid r_a = r) = P(b \text{ does not transmit in the same slot}) \cdot$$
$$P(\text{no transmission from Reg 1 and 2 during a mini-slot} \mid r_a = r) \cdot$$
$$P(\text{no transmission from Reg 3 during } 2T+1 \text{ mini-slots} \mid r_a = r)$$

$$= (1-p) \cdot e^{-pP(I_N|G(1))\bar{N}(G(1))} \cdot e^{-pP(I_N|G(2))\bar{N}(G(2))} \cdot e^{-(2T+1)pP(I_N|G(3))\bar{N}(G(3))}, \quad (36)$$

where $P(I_N|G(g))$ is given by (22), and $\bar{N}(G(g))$ represent the expected number of vehicles in Region $g$. Hence, the local throughput and expected progress for the transmitter node located at $a$ are given by

$$\rho_a = (1 - e^{-\bar{N}^a_{a-R}}) \frac{p}{\tau} \int_0^R P(a \to b \mid r_a = r) dF_{r_a}(r) \quad (37)$$

$$\pi_a = (1 - e^{-\bar{N}^a_{a-R}}) \frac{p}{\tau} \int_0^R P(a \to b \mid r_a = r) r dF_{r_a}(r). \quad (38)$$

The average local throughput and progress for the whole network can be found accordingly by substituting (37) and (38) into (35), respectively. In the next section, we aim to optimize the expected progress $\pi$ by finding the optimal transmission probability $p$, and the corresponding optimal sensing probability.

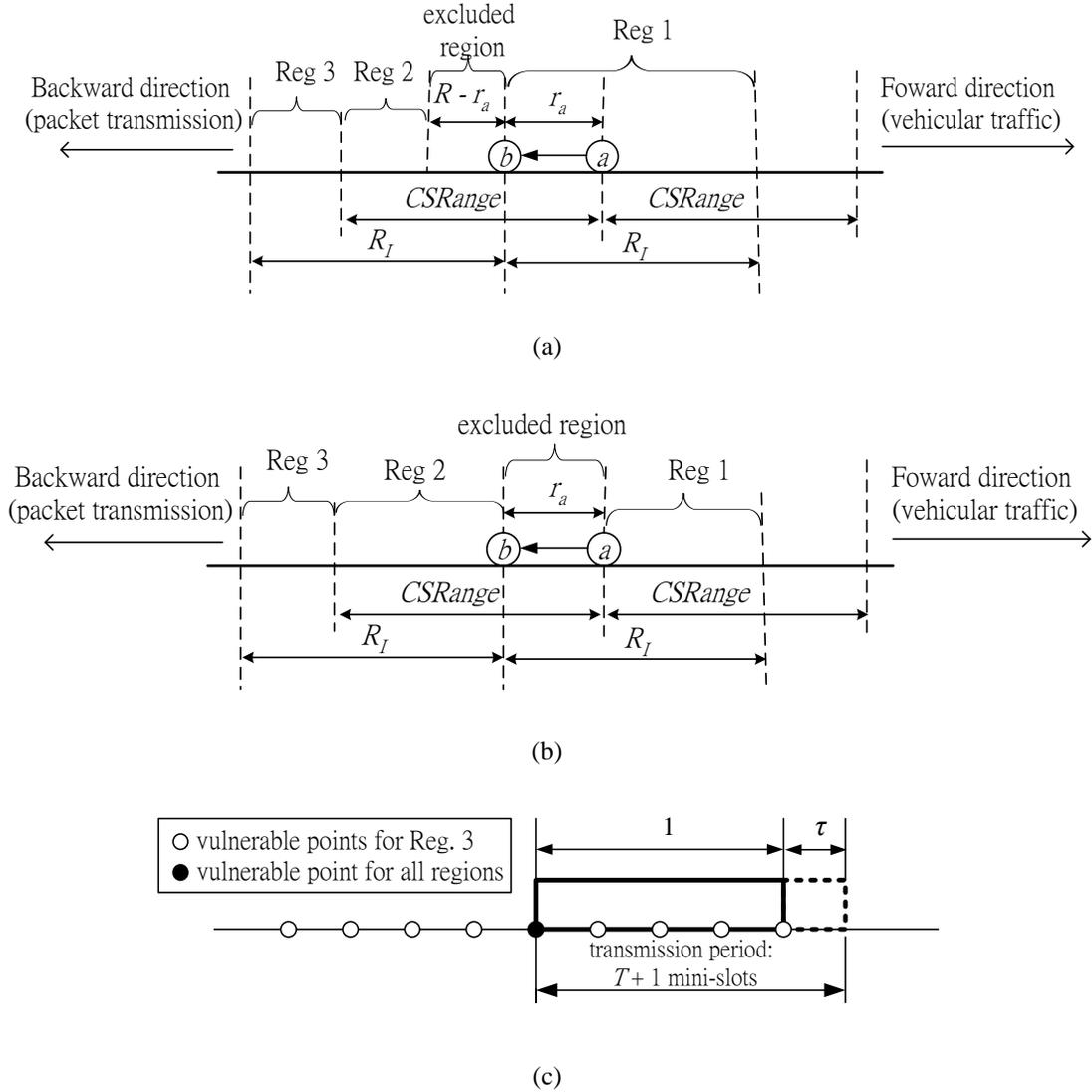

Figure 6. a) Regions that a neighbor of node *b* could be located in under the MPR and MP strategies; b) under the NP strategy; and c) the period for the transmission from *a* to *b* vulnerable to transmissions from different regions.

## VI. NUMERICAL RESULTS

To validate our analytical model, we simulate a road segment of length 5 km. We assume that there are no cars joining and leaving the route at junctions, cars only arrive at location 0 at a constant rate (denoted by $\alpha$). Every vehicle moves along the road with respect to a velocity profile, $v(x)$ as shown in Figure 7, which describes a slowdown from positions 1 to 3 km.

Intuitively, we know that the slowdown region will result in higher node density than the other parts of the road, where the mean density profile $n(x)$ can be computed by the fluid model. In the simulations, we have $R = 0.1$ km, $\beta = 10$, $\gamma = 4$. For CSMA, $\tau = 0.25$ and $CSRange = 0.178$.

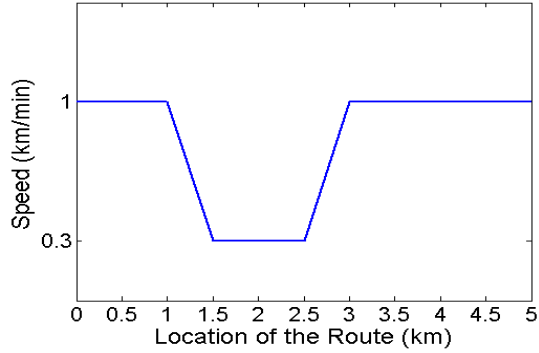

Figure 7. Velocity profile used for examining the non-homogeneous density case.

From the analytical results, the optimal transmission probabilities $p$ for slotted ALOHA and CSMA are sought to maximize the average progress $\pi$ in the network, which are plotted in Figure 8 against the arrival rate of vehicles to the road segment. We can see from Figure 8a that the NP strategy has the largest optimal transmission probability that maximize the average progress, followed by the MP strategy for the slotted ALOHA scheme. This is because NP aims to reduce the amount of interference in the network as much as possible, nodes should transmit more frequently in order to achieve optimal performance when there are less amount of interference. In Figure 8b, we can see that CSMA has a smaller optimal transmission probability than slotted ALOHA, because nodes suppress their transmissions to avoid collisions when the channel is sensed busy. The optimal transmission probabilities $p$ found for slotted ALOHA and CSMA are served as inputs to the simulations. For CSMA, the corresponding optimal sensing probability $p'(x)$ is found based on the optimal $p$ and density profile $n(x)$, the reader is referred to Appendix I for details.

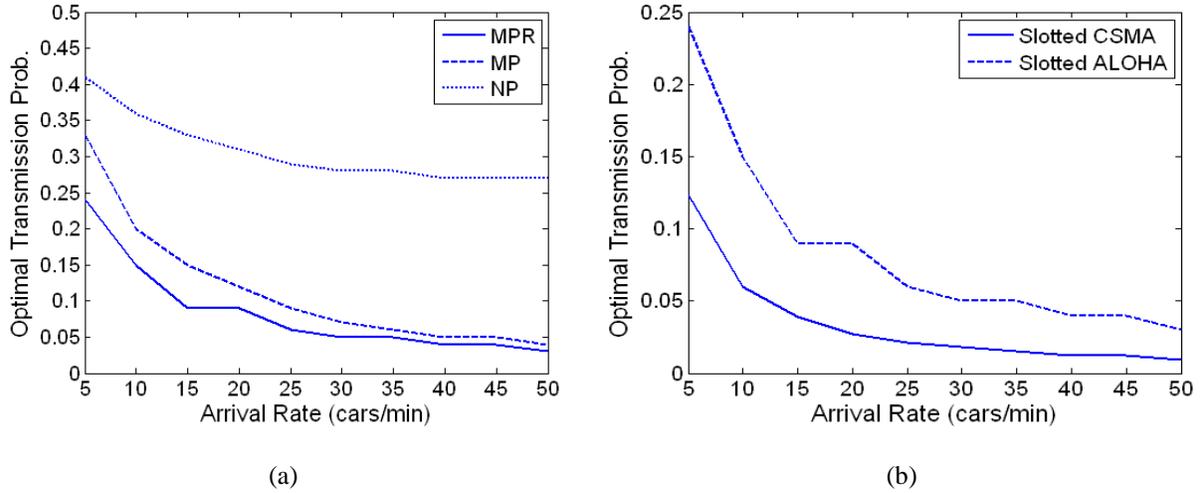

(a) (b)

Figure 8. Transmission probability that maximize the average progress in a) slotted ALOHA; and b) slotted ALOHA and CSMA for the MPR strategy.

In the simulation, we first identify the number of nodes in the road segment at a time instance, denote it by $N$. For slotted ALOHA, each node in the network transmits with probability $p$ in every slot. For each transmission from a transmitter to its next relay, we examine every node in the road segment to see if it is covered by the interference range of this transmission. We identify those nodes which are not covered by any interference ranges of others. If the identified node is a relay, a transmission to it is said to be successful and the progress of this transmission is recorded. Let $N_T$ be the number of nodes with successful transmissions, the single-hop throughput $\rho$ is computed as $N_T/N$. Let $D$ be the sum of the recorded progresses, the average progress per slot of a node, $\pi$, is computed as $D/N$.

While for slotted CSMA, each node in the network senses the channel with probability $p'$ in every mini-slot. A transmission is successful only if the link is not being interfered during the transmission period of $T + 1$ mini-slots. The number of successful transmissions made during a mini-slot duration is identified, let us denote it as $M_T$, and the corresponding progresses of successful transmissions are recorded. The single-hop throughput $\rho$ is computed as $M_T/(\tau N)$. The sum of the recorded progresses is denoted as *D*, and the average progress per slot of a node, $\pi$, is computed as $D/(\tau N)$.

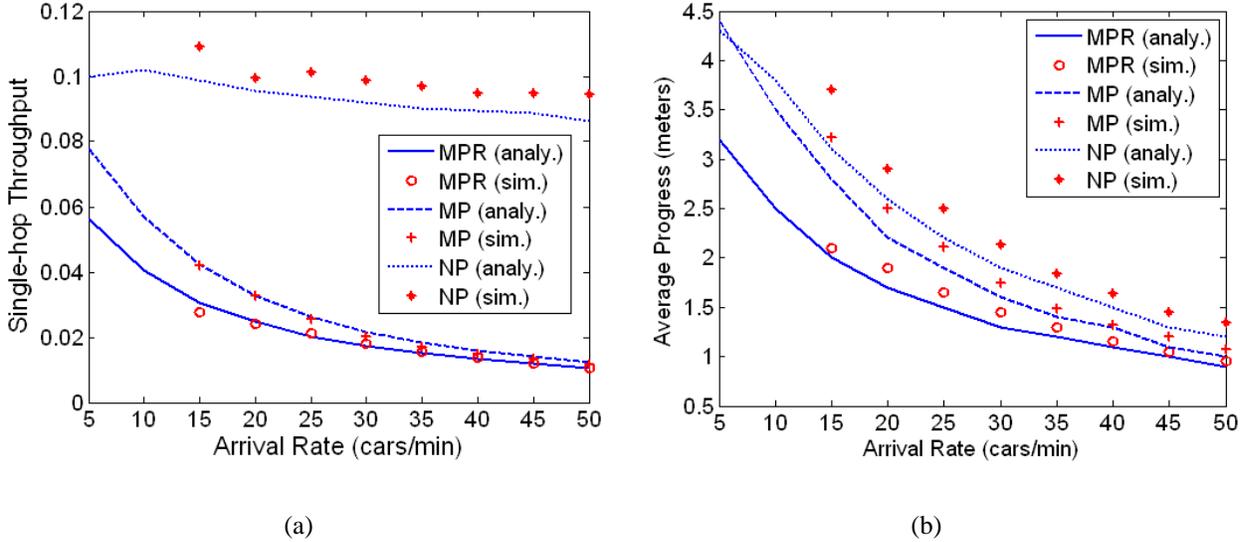

(a)                      (b)

Figure 9. a) Single-hop throughput; and b) average progress against vehicle arrival rate in slotted ALOHA for the MPR, MP, and NP strategies.

Figure 9b plots the analytical and simulated maximized average progress as a function of the vehicle arrival rate of the three routing strategies in slotted ALOHA network, the corresponding single-hop throughput is shown in Figure 9a. Each simulated data is obtained by averaging the results of 500 simulation runs. In general, we can find that the analytical and simulated results have a pretty good match. According to the figures, the network performance degrades as the arrival rate (or density) increases. This is due to the fact that the amount of network traffic increases as node density grows, which will lead to more interference. However, we can see that the single-hop throughput of the NP strategy remains stable and decreases very slowly as the arrival rate increases, which gives rise to the largest single-hop throughput among the three strategies considered. This is because transmitters choose the nearest backward neighbor as the relay node in the NP strategy, transmitters will tend to choose nearer neighbor for transmission as node density increases. Hence, interference ranges of transmissions become smaller since we assume nodes will use the minimum required power for transmission in the NP strategy. This balances off the performance degradation caused by the increasing traffic load (or amount of interference) due to higher node density. As a result, the single-hop throughput of the NP strategy is fairly the same when the arrival rate is high (e.g., about 0.09).

On the other hand, for the MPR and MP strategies, the single-hop throughput drops as the arrival rate increases. In MP, a transmitter chooses the node with the most progress as its intended receiver. As node density increases, the node will tend to choose a more distanced node as relay, which results in larger interference range and thus degradation of single-hop throughput. For MPR, the transmission range is always equal to $R$, which yields the greatest possible interference range, and more nodes will be interfered as the network traffic load becomes larger. Therefore, MPR has the worst performance among the three strategies.

It is also interesting to note that the NP strategy also yields the best progress performance among the three strategies. This gives us a key message that minimizing the interference of transmissions is of higher priority than maximizing the per hop progress for achieving optimal progress performance in a linear vehicular network. With reference to Figure 8a again, the optimal transmission probability for the NP strategy is around 0.27 when node density is high. Such property is significant for network design and planning in VANET, where network topology keeps changing (for example, given that node density is high enough, we can have each node just transmits to its nearest neighbor with a transmission probability $p$ of 0.27, regardless of how the topology changes).

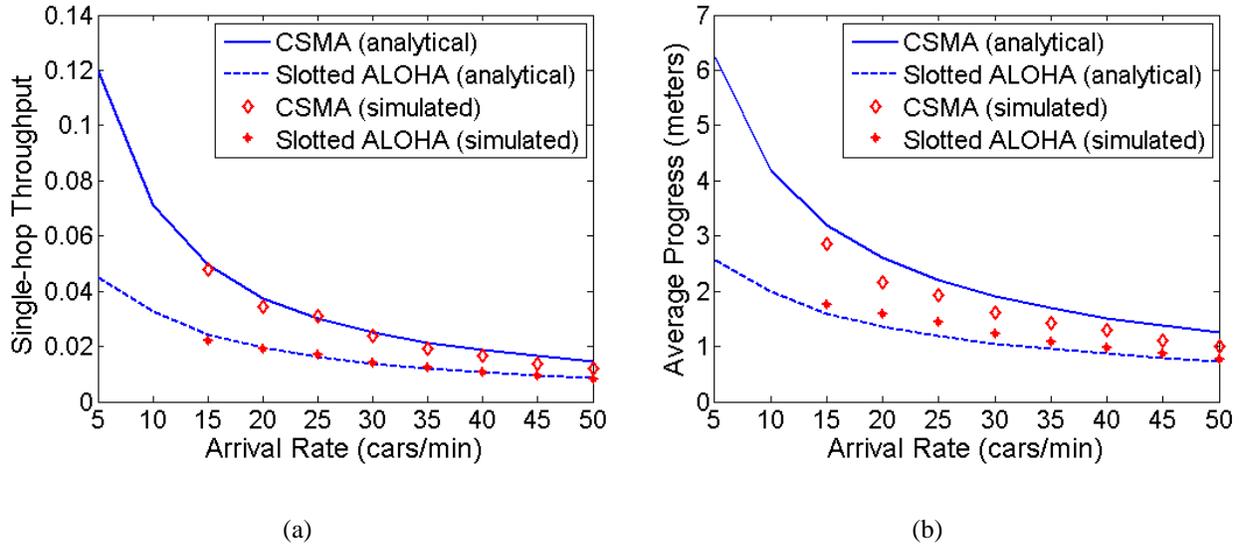

(a)    (b)

Figure 10. a) Single-hop throughput; and b) average progress against vehicle arrival rate in slotted ALOHA and CSMA for the MPR strategy.

To evaluate the analysis for CSMA protocol, the analytical and simulated results of the maximized average progress $\pi$, and the corresponding single-hop throughput $\rho$ for the MPR strategy in slotted ALOHA and CSMA are plotted in Figure 10b and Figure 10a respectively as a function of the arrival rate of cars to the road (for proper comparison, the average progress and single-hop throughput of slotted ALOHA is divided by $1 + \tau$ to include the propagation time in a slot). We can see from the figures that CSMA outperforms slotted ALOHA as expected due to the carrier-sensing mechanism that avoid packet collisions, and the analytical results match closely with the simulated results, which confirms our analysis.

Instead of maximizing the average local throughput $\pi$, and progress $\rho$ for the whole network, we can optimize $\pi_x$ and $\rho_x$ for a specific location point $x$ in the road segment. The optimal transmission probabilities $p(x)$ for slotted ALOHA and CSMA (and the corresponding optimal sensing probability found according to (41) in Appendix I) are shown in Figure 11a and Figure 11b respectively. From the figures, we can see that region with higher node density (the slowdown region from 1 to 3 km) results in lower optimal transmission and sensing probabilities, which is intuitively correct.

In practice, velocity and flow data measured by inductive loop detectors can be gathered by road-side infrastructure nodes that are connected to the wired backhaul network. The road-side nodes can then compute the optimal transmission probability profile that is a function of space and time based on our analytical results, and broadcast it to vehicles that enter the corresponding road regions. The results in this paper can also serve as a fundamental building block for the design of VANET protocols with adaptive transmission/sensing rate according to the vehicular density in the future.

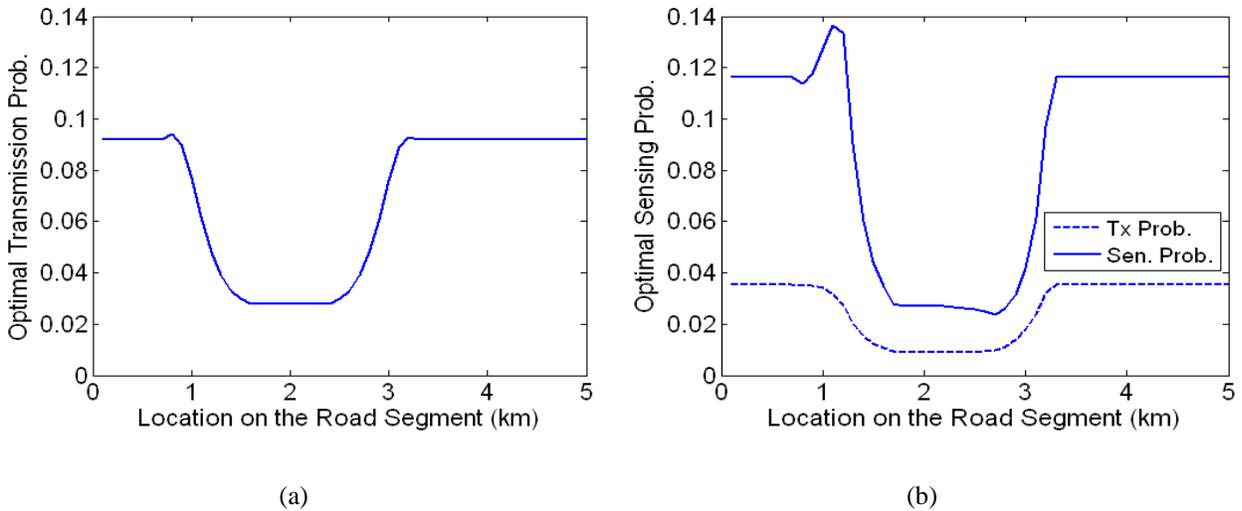

(a) (b)

Figure 11. a) Optimal transmission probability for slotted ALOHA; and b) Optimal sensing probability for CSMA as a function of the Location Space for the MPR strategy.

## VII. CONCLUSION

Vehicles in urban road networks do not distribute homogeneously as generic nodes do in mobile ad-hoc networks. However, most of the existing studies on communication protocol performance have the implicit assumption that nodes are distributed homogeneously in the network, which is inappropriate in VANETs and may lead to unreliable results.

In this paper, we propose a novel methodology to analyze protocol performance in VANETs with practical vehicle distribution in urban environment. To our knowledge, this is the first attempt in the literature that explicitly addresses the heterogeneity of node distribution in VANETs. The proposed approach employs the stochastic traffic model for computing the vehicular density dynamics in urban road networks from a velocity profile, which can be constructed in practice based on data collected by navigation systems that are widely installed in vehicles nowadays. Based on the density knowledge, we have shown with illustrative examples that the throughput and progress performance of different channel access protocols with different routing strategies can be derived. We have also confirmed the accuracy of our analysis through extensive simulations, and demonstrated that the optimal transmission probability (or sensing probability) profile as a function of the location space can be identified through the analytical model, which is significant for system engineering and network planning in VANETs.

The methodology presented in this paper is applicable to more elaborated urban traffic models. For example, routes with arrival and departure of vehicles at road junctions, multiple lanes, bi-directional traffic. More complicated urban road network (e.g., two-dimensional road topology) can be represented by superposing multiple versions of urban routes. Furthermore, we can consider more practical velocity profile. For instance, velocity profile that is a function of vehicular density to approximate interactions between vehicles [9]. In general, the throughput and progress performance of other routing strategies and channel access protocols can be evaluated similarly with the approaches presented in this paper, and optimization of communication protocols with adaptive transmission/sensing rate based on the vehicular density dynamics can be investigated in the future.

**Appendix I: Determination of the Sensing Probability for CSMA**

In this appendix, we determine the relation between the transmission probability $p$ and channel-sensing probability $p'$ for slotted non-persistent CSMA. First, we assume that

$$p = p'P_i, \tag{39}$$

where $P_i$ is the probability that the channel is sensed idle. That is, a node transmits in a slot only if it senses the channel and the channel is sensed idle.

Let us consider a node located at position $x$, we assume the probability that a slot is sensed idle is $exp(-pN_{CS}(x))$, where $N_{CS}(x) = \bar{N}_{x-CSRange}^{x+CSRange}$ is the expected number of nodes within the carrier-sensing range of the considered node, which can be found from vehicular density profile $n(x)$. Hence, the expected value of the idle period $I$ is

$$\bar{I} = \sum_{n=1}^{\infty} n\tau e^{-npN_{CS}(x)}\left(1 - e^{-pN_{CS}(x)}\right)$$

$$= \sum_{n=1}^{\infty} n\tau e^{-npN_{CS}(x)} - n\tau e^{-(n+1)pN_{CS}(x)}$$

$$= \sum_{n=1}^{\infty} \tau e^{-npN_{CS}(x)} = \frac{\tau e^{-pN_{CS}(x)}}{1 - e^{-pN_{CS}(x)}}.$$

Since the transmission period is $1 + \tau$, thus,

$$P_i = \frac{\bar{I}}{\bar{I} + 1 + \tau} = \frac{\tau e^{-pN_{CS}(x)}}{1 + \tau - e^{-pN_{CS}(x)}}. \tag{40}$$

By substituting (40) into (39), we obtain an equation that express $p'$ in terms of $p$, $\tau$, and $N_{CS}(x)$ as a function of the location space $x$ as

$$p'(x) = \frac{(1 + \tau - e^{-pN_{CS}(x)})p}{\tau e^{-pN_{CS}(x)}}. \tag{41}$$